\documentclass[prl,twocolumn,showpacs,floatfix]{revtex4-1}
\usepackage{graphicx}
\usepackage{array}
\usepackage{amsmath,amsfonts,amssymb}
\usepackage[ansinew]{inputenc}
\usepackage{color}

\newcommand{\Figref}[1]{Fig.~\ref{#1}}

\begin{document}
\title{Electroluminescence from a polythiophene molecular wire suspended in a plasmonic scanning tunneling microscope junction.}

\author
{Gaël Reecht$^1$, Fabrice Scheurer$^1$, Virginie Speisser$^1$, Yannick J. Dappe$^{2}$, Fabrice Mathevet$^3$,\\
Guillaume Schull$^1$}

\altaffiliation{guillaume.schull@ipcms.unistra.fr}
\affiliation{$^1$Institut de Physique et Chimie des Mat\'eriaux de Strasbourg, UMR 7504 (CNRS -- Universit\'e de Strasbourg), 67034 Strasbourg, France\\ 
$^2$SPEC (CNRS URA2464), SPCSI, IRAMIS, CEA Saclay, 91191 Gif-Sur-Yvette, France\\
$^3$Laboratoire de Chimie des Polym\`eres,
UMR 7610 (CNRS -- Universit\'e Pierre et Marie Curie), 75252 Paris, France}

\begin{abstract}
The electroluminescence of a polythiophene wire suspended between two metallic electrodes is probed using a scanning tunneling microscope. Under positive sample voltage, the spectral and voltage dependencies of the emitted light are consistent with the fluorescence of the wire junction mediated by localized plasmons. This emission is strongly attenuated for the opposite polarity. Both emission mechanism and polarity dependence are similar to what occurs in organic light emitting diodes (OLED) but at the level of a single molecular wire.

\end{abstract}

\date{\today}


\maketitle

Controlling the luminescence of a single molecule directly bridging metallic electrodes is a challenging key issue towards molecular optoelectronics \cite{Galperin2012}. Recent experiments based on non-imaging methods have shown that the electroluminescence of complex molecular-nanotube \cite{Marquardt2010} and metallic 
nano-cluster \cite{Lee2003} junctions can be excited. Reaching a deeper understanding of these mechanisms requires atomic scale control of the geometrical parameters of the molecular junction while simultaneously monitoring electronic and optical characteristics \cite{Berndt2010}. In a pioneering experiment \cite{Qiu2003}, a scanning tunnelling microscope (STM) was used to excite with atomic-scale accuracy the fluorescence of a single molecule separated from the electrodes by thin insulating layers. While such a \textit{weak electrode-molecule coupling} is highly favourable for the observation of molecular fluorescence \cite{Qiu2003,Dong2004,ifmmode2005,Kabakchiev2010,Chen2010}, a \textit{direct connection} to the metallic leads is desirable towards integration into molecular scale circuits. To this end, the light emission from atomic \cite{Schull2009,Schneider2010} and molecular contacts \cite{Schneider2012} was recently probed, revealing the influence of the current shot noise at elevated conductances. In such highly coupled cases, however, luminescence mechanisms intrinsic to the molecule are quenched because of the strong hybridization with the electrode states \cite{Avouris1984,Hoffmann2002,Rossel2010}.

Here we use the tip of a scanning tunnelling microscope to controllably lift a unique $\pi$-conjugated polymer chain from a Au(111) surface  \cite{Lafferentz2009}. In this configuration, both extremities of the $\pi$-conjugated wire are directly connected to the electrodes, whereas a long part of the polymer chain is suspended in the junction. Under positive sample bias, light is emitted at the wire junction. Optical spectra reveal a broad resonance whose maximum does not shift with voltage. Based on a simple model, and in agreement with predictions \cite{Buker2002,Galperin2006}, this emission is traced back to the recombination of electrons injected from the tip in the lowest unoccupied molecular orbital (LUMO) with holes injected from the sample in the highest occupied molecular orbital (HOMO) of the wire junction, showing that direct molecule--electrode coupling and fluorescence can be associated in a single molecular junction. For the opposite polarity, the photon intensity is strongly reduced (quantum yield attenuated by a factor 10$^{3}$). Our model suggests that this behaviour is due to the non-centred HOMO--LUMO gap respectively to the Fermi level of the electrodes at zero voltage, and from the slight asymmetry between the wire/tip and wire/substrate coupling strength. Both emission mechanism and polarity dependence are consistent with the behaviour of a single polymer light emitting diode. Finally, surface plasmons localised at the tip-sample cavity, which are known to amplify inelastic electronic transitions in noble metal STM junctions \cite{Rossel2010}, reveal some unexpected interactions with the wire optoelectronic properties.

The experiments were performed with an Omicron STM operating at 4.6\,K in ultrahigh vacuum. The setup was adapted to light emission measurements following the design developed in \cite{Keizer2009}. In this setup a lens (f-number = 1.5) is fixed to the STM head so that the tip-sample junction can be precisely located at the lens focal point. The collimated light is redirected through an optical viewport outside of the vacuum chamber and is refocused on an optical fibre bundle. The fibre bundle is connected to a detection unit composed of a grating spectrograph (Princeton Instruments Acton Series SP-2300i) and a liquid nitrogen cooled CCD camera (Princeton instruments PyLoN-100BR-eXcelon). An overall 1.5\% detection efficiency is estimated at 500\,nm. Au(111) samples and etched W tips were sputtered with argon ions and annealed. To increase their plasmonic response, the STM-tips were covered with gold by micrometer-scale indentations into the sample. The on-surface synthesis of polythiophene nanowires is detailed in \cite{Reecht2013}. Differential conductance spectra were acquired for a fixed tip-sample distance using lock-in detection (RMS amplitude = 10\,mV, modulation frequency = 740\,Hz).

\begin{figure}
  \includegraphics[width=.95\linewidth]{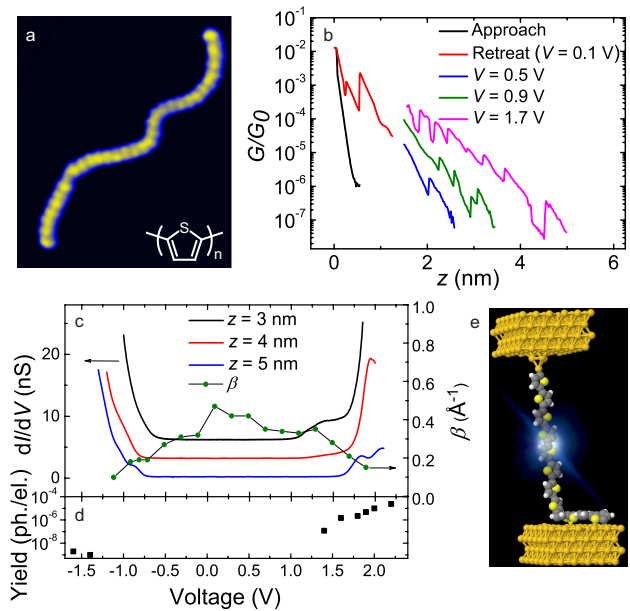}
  \caption{\label{fig1} (a) STM image (9.0 $\times$ 10.8 nm$^2$, $I$ = 2 nA, $V$ = 0.1 V) of a polythiophene wire (inset) polymerized on a Au(111) sample. (b) Normalized conductance $G/G_0$ vs tip-sample distance $z$ for a polythiophene wire suspended in the junction for different voltages. The black curve corresponds to the initial approach of the clean STM tip to a wire extremity. The point of contact defines the origin of the abscissa. (c) Conductance d$I$/d$V$ spectra (lines) acquired at different tip--sample distances and inverse decay length $\beta$ (circles) as a function of $V$, for a given suspended wire. The spectra acquired at $z$ = 4 nm and $z$ = 3 nm are offseted by 3 and 6 nS respectively. (d) Light emission efficiency (squares) as a function of $V$. (e) Artistic view of a fluorescent polythiophene junction.} 
\end{figure}       

We start by discussing the procedure used to suspend an individual polythiophene wire (\Figref{fig1}a) in the junction. The STM tip is first located atop one extremity of the polymer deposited on a Au(111) surface \cite{Reecht2013}, then approached to the wire up to the formation of a contact, and retracted to its initial position while constantly recording the current traversing the junction. The success of the procedure is attested by the substantially changed slope upon retraction (\Figref{fig1}b). Periodic current ''jumps'' occuring during the lifting procedure are associated to the successive detachments of thiophene units from the surface. Despite these abrupt conductance changes the overall conductance traces can be fitted as $G(z)$ $\propto$ $G_c exp(-\beta z)$, where $G_c$ is the conductance at contact, $z$ is the tip--sample distance, and $\beta$ reflects the  ability of the wire to transport current \cite{Koch2012}. For a sample voltage $V$ = 0.1 V, we find $\beta = 0.40 \pm 0.08$ \AA$^{-1}$, in good agreement with predictions \cite{Magoga1997}. At higher voltages a lowering of $\beta$ is observed (\Figref{fig1}b). d$I$/d$V$ spectra recorded for different suspended polymer lengths (\Figref{fig1}c) reveal a first resonance at $V\approx$ -0.8 V, another at $\approx$ 1.25 V which progressively disappears when the tip is retracted because of the weaker current (see also supplementary section S1 \cite{supp}), and a more intense resonance around 2 V. At this stage it is impossible to assign a LUMO or HOMO origin to these resonances. Indeed, both extremities of the wire are directly coupled to the electrodes (\Figref{fig1}e) and the voltage partially drops at each interface (assuming no voltage drop along the wire itself \cite{Cuevas2010}) causing a displacement of the HOMO and LUMO states in the d$I$/d$V$ spectra respectively to their positions at $V =$ 0.
As reported for graphene nanoribbons \cite{Koch2012}, we can see that the reduction of $\beta$ with $V$ occurs when the voltage reaches a d$I$/d$V$ resonance (dots in \Figref{fig1}c). This effect results from the cross-over from non-resonant to resonant transport regime \cite{Mujica1994,Koch2012}.

\begin{figure}
  \includegraphics[width=1.00\linewidth]{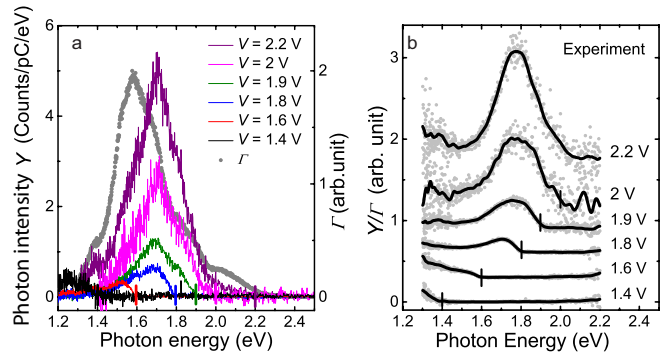}
  \caption{\label{fig2} (a) Raw and (b) plasmon--corrected spectra of the light emitted by a polythiophene wire suspended in the STM junction for different voltages (1 nA $< I <$ 5 nA). To prevent damages to the wire due to an increased current at elevated voltages, the spectra were acquired at $z$ = 3 nm ($V$ = 1.4 V and $V$ = 1.6 V), $z$ = 4 nm ($V$ = 1.8 V and $V$ = 1.9 V) and $z$ = 5 nm ($V$ = 2 V and $V$ = 2.2 V). The corresponding plasmon amplification function $\mathnormal{\Gamma}(h\nu)$ is shown (dots) in (a). The spectrum at $V$ = 1.4 V was scaled by a factor 5 in (a). As expected for electroluminescent processes \cite{Schull2009}, the energy of the photons does not exceed the energy of the electrons ($h\nu$ $<$ $eV$), explaining the cut-off (vertical dashes) at the high-energy edge of the spectra.} 
\end{figure}

We now turn to the electroluminescent properties of the suspended polymer. In figure \ref{fig2}a we display optical spectra acquired at different voltages for the molecular wire probed in \Figref{fig1}c. For $V$ = 1.4 eV, the quantum yield of the emission process is in the order of 10$^{-7}$photon/electron and gradually rises to $\approx$ 10$^{-5}$ at higher voltage (\Figref{fig1}d) due to the larger emission bandwidth. For the opposite polarity ($V$ = -1.4 V and $V$ = -1.6 V) and the same tip-sample distances, an extremely weak emission ($\approx$ 10$^{-9}$photon/electron) is detected. As demonstrated below, this strong variation of the emission efficiency with polarity indicates that the wire states are involved in the luminescence process.

\begin{figure}
  \includegraphics[width=1.00\linewidth]{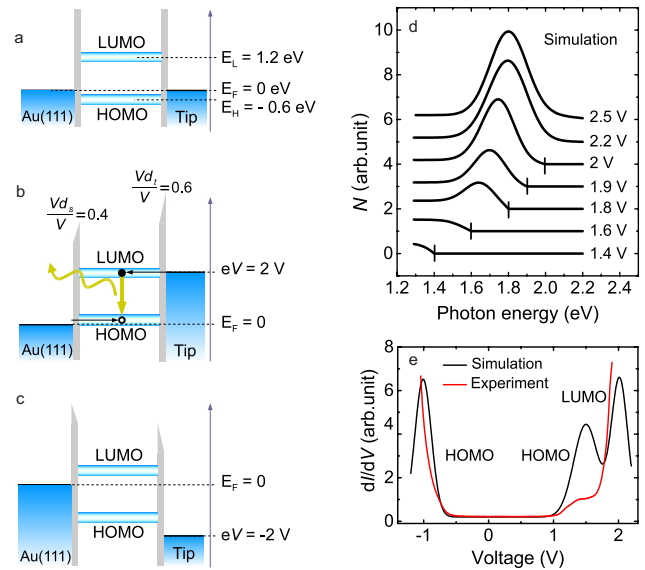}
  \caption{\label{fig3} (a) Sketch of the polymer junction representing the energies of the HOMO ($E_H$) and LUMO ($E_L$) states at zero voltage.  For sufficiently high positive sample voltage (b) the HOMO is above the Fermi level of the sample (taken as reference) and the LUMO below the Fermi level of the tip. In this condition, both states can be either occupied or unoccupied at a given time, and electrons injected from the tip in the LUMO can radiatively decay in the partially emptied HOMO. For the opposite polarity (c) the LUMO remains above the Fermi level of the sample and no intra-molecular radiative transition occurs. (d) Simulations of the light emission spectra as a function of the applied voltage. (e) Comparison between simulated (black line) and experimental (red line) d$I$/d$V$ spectra for a suspended wire.} 
\end{figure}

In STM-induced light emission from noble metal electrodes, plasmons localized at the tip-sample junction \cite{Rossel2010} strongly amplify any radiative transition. The energy dependence of this amplification ($\mathnormal{\Gamma}(h\nu)$ in \Figref{fig2}a) can be deduced from optical spectra of the pristine junction (supplementary section S2 \cite{supp}). Assuming that $\mathnormal{\Gamma}(h\nu)$ does not significantly change with tip-sample distance (see supplementary section S3 \cite{supp}), the impact of the plasmons on the shape of the spectra can be corrected by normalizing each spectrum by $\mathnormal{\Gamma}(h\nu)$ \cite{Schneider2011}. For the polythiophene junctions, the corrected spectra (\Figref{fig2}b) reveal the progressive apparition with voltage of a broad resonance centred at 1.8 eV. Interestingly, this resonance does not shift with voltage for $V >$ 2 V. Mechanisms where electrons decay from the wire states directly to states of the electrodes, or \textit{vice versa}, cannot account for this observation (see supplementary section S4 \cite{supp}). This observation rather suggests an emission associated to electrons decaying from one molecular state to another \cite{Qiu2003}. To confirm this assertion we build up a two Gaussian level model whose parameters at $V$= 0, given in figure \ref{fig3}a, are justified below. In figure \ref{fig3}b and c (see also supplementary section S5 \cite{supp}), we see how the electronic properties of this junction evolve as a function of voltage. For a high enough positive voltage (\Figref{fig3}b), and an appropriate voltage drop at the wire--substrate ($V\!d_s$) and wire--tip ($V\!d_t$) interfaces, electrons injected in the LUMO may recombine with holes injected in the HOMO \cite{note2}. In a simple approximation where the number of emitted photons ($N$) scales with the available inelastic transitions between two partially occupied states (case of \Figref{fig3}b), the spectral and voltage dependencies of this emission follow
\begin{equation}
\label{eq:fluo}
{N(h\nu, eV)} \propto \int_{h\nu}^{eV} f_{L}(E) f_{H}(E-h\nu) dE,   \; \ \ \ \text{for} \ \textit{V} > 0,
\end{equation}      
\noindent                             
where $f_{L}$ and $f_{H}$ are Gaussian functions representing the LUMO and HOMO states
\begin{equation}
\label{eq:Lumo}
{f_{i}(E)} = \frac{1}{\sigma\sqrt{2\pi}}\exp{-\frac{(E - (E_i+e\,V\!d_s))^{2}}{2\sigma^{2}}}, \;  i = H\; \text{or}\; L ,
\end{equation}     
\noindent
where $\sigma = \frac{W}{2\sqrt{2\ln{2}}}$, $W$ = 0.155 eV is the width of the molecular orbitals as deduced from the experimental d$I$/d$V$ spectra (\Figref{fig1}c), and $E_i$ the energy of the state $i$ at $V$ = 0. For $V\!d_t$/$V\!d_s$ = 1.5, $E_H$ = -0.6 eV and $E_L$ = 1.2 eV, the simulated optical spectra (\Figref{fig3}d) are in very good agreement with the experimental data (\Figref{fig2}b). Our model further explains the absence of emission at negative voltages (\Figref{fig3}c). Indeed, in this case a high negative voltage ($\approx$ -3 V) is required to shift the LUMO below the Fermi level of the sample (supplementary section S5 \cite{supp}).
For sake of consistency, we compare for the same parameters a calculated (see supplementary section S6 \cite{supp}) and an experimental d$I$/d$V$ spectrum acquired with the same suspended wire (\Figref{fig3}e). The good match between these spectra enables the assignment of the HOMO and LUMO orbitals (\Figref{fig3}e) and further validates the electrofluorescence model.  These parameters are also consistent with simulations based on density functional theory (see supplementary section S7 \cite{supp}) which support our interpretation. As theoretically predicted \cite{Buker2002,Galperin2006}, these results show that the asymmetric position of the HOMO--LUMO gap with respect to $E_F$ at $V$ = 0 and the asymmetric voltage drop repartition at the interfaces are decisive for the emission properties of the junction.  

At a first glance, our model looks similar to what has been reported for double tunneling junction experiments where the molecular emitter is separated from the two electrodes by thin insulating layers  \cite{Qiu2003,ifmmode2005,Chen2010}. Indeed, in both cases the voltage drops partially on each side of the molecule which provides the adapted energy configuration (\Figref{fig3}b) for fluorescence. However, the direct connection between the molecule and the electrodes in our experiment broadens the orbitals of the conjugated wire. This explains why the fluorescence spectra (\Figref{fig2}b) exhibit broad features rather than sharp resonances. In that sense, the registered optical spectra are characteristic of the overall junction (i.e. wire and contacts) and not of the isolated wire.\\
 
\begin{figure}
  \includegraphics[width=.95\linewidth]{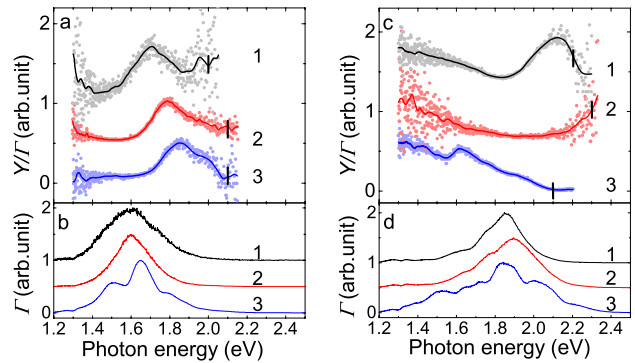}
  \caption{\label{fig4}(a) Plasmon--corrected light emission spectra of different wire junctions ($I$ = 1.5 nA, 0.5 nA and 0.5 nA for spectra 1, 2 and 3 respectively) and (b) their respective plasmon amplification functions $\mathnormal{\Gamma}(h\nu)$ for non--resonant plasmon--emitter conditions. The $\mathnormal{\Gamma}(h\nu)$ spectra all show a resonance around 1.6 eV, excluding an influence of the plasmons on the shifts observed in (a). The data in (c) ($I$ = 1.5 nA, 1 nA and 1 nA for spectra 1, 2 and 3 respectively) and (d) pertain to other set of measurements where resonant plasmon--emitter conditions were observed. The vertical dashes mark the e$V$ = h$\nu$ quantum cut-off.} 
\end{figure}

Other polythiophene wires reveal equivalent emission properties (\Figref{fig4}a), except for a slight shift of the fluorescence maximum (1.85 eV $\pm$ 0.15 eV). Similar shifts are reported in fluorescence spectra of polythiophene in solution or in thin film (from 1.6 eV to 2.2 eV) \cite{Rumbles1996,Bolognesi2000,Korovyanko2001}. Conformational changes of the wires are known to impact the electron delocalization length and consequently the emission wavelength \cite{Xu1993}. We speculate that this is also the case in our experimental configuration, where the suspended wires may adopt different conformations (e.g. tilting of bases) depending on the details of the lifting procedure. Experimental data provided in supplementary section S8 \cite{supp} corroborate this interpretation. However, distinct emission properties are observed (\Figref{fig4}c) when the plasmon resonance maximum is centred at the same energy than the wire fluorescence (resonant condition). In this case, the plasmon-corrected spectrum reveals either no, strongly blue or red shifted resonances. This striking behaviour suggests that, for resonant conditions, more complex interaction mechanisms occur between the plasmon modes and the wires states. Strong coupling between the emitter and plasmon modes \cite{Salomon2009} as well as plasmon assisted excitations of the emitter \cite{Dong2010} are possible explanations. 

Finally, it is interesting to discuss the absolute efficiency of the overall emission process ($Q$ $\approx$ 2.5 $\times$ $10^{-5}$ photon/electron at $V$ = 2.2 V) in the scope of the linewidth of the light emission peak ($W_f$ $\approx$ 0.25 eV at $V$ = 2.2 V). From this last parameter we can deduce the lifetime of the excited state ($\tau_{ex}$ = $\hbar$/$W_f$ $\approx$ 2.6 fs) which depends on all the possible radiative and non-radiative des-excitation channels of the molecular junction \cite{Wu2008}. As deduced from the low emission efficiency, the lifetime of the non-radiative decay channels ($\tau_{nonrad}$) is much shorter than the radiative ones ($\tau_{rad}$), and $\tau_{ex}$ $\approx$ $\tau_{nonrad}$. Since the emission efficiency verifies $Q$ $\approx$ $\tau_{nonrad}$/$\tau_{rad}$, we can estimate the lifetime for the junction emission $\tau_{rad}$ $\approx$ 100 ps. This value corresponds to the ratio between the intrinsic fluorescence lifetime of the wire junction and an amplification factor linked to localized plasmons  \cite{Wu2008}. Overall, $\tau_{rad}$ is approximately an order of magnitude shorter than the radiative lifetime of polythiophene in solution \cite{Rumbles1996}, suggesting a plasmonic amplification of $\approx$ 10 \cite{note3}. 

Our results shed light on the electroluminescent properties of single molecules bridging metallic junctions. They confirm previous theoretical works \cite{Buker2002,Galperin2006} which predict that appropriate molecule-electrode connections may allow for intramolecular radiative transitions. We demonstrate that these transitions are directly affected by the wire conformation, the interface-induced broadening of the molecular states and plasmon modes localized at the STM junction. In the scope of future applications, the polarity dependence of the emission is another remarkable property of the presented wire junctions. This strong similarity with OLEDs paves the way towards single molecular optoelectronic components. 

\noindent
The authors thank Laurent Limot for stimulating discussions, and Jean-Georges Faullumel, Michelangelo Romeo and Olivier Cregut for technical support. The Agence National  de  la  Recherche, contract TRANSMOL ANR-2010-JCJC-1004, the Région Alsace, and the International Center for Frontier Research in Chemistry (FRC) are acknowledged for financial  support.

\bibliographystyle{prsty}
\bibliography{lightwirenat}

\begin{thebibliography}{10}

\bibitem{Galperin2012}
M. Galperin and A. Nitzan, Phys. Chem. Chem. Phys. {\bf 14},  9421  (2012).

\bibitem{Marquardt2010}
C.~W. Marquardt, S. Grunder, A. Blaszczyk, S. Dehm, F. Hennrich, H.~v.
  Lohneysen, M. Mayor, and R. Krupke, Nat. Nanotechnol. {\bf 5},  863–867
  (2010).

\bibitem{Lee2003}
T.-H. Lee and R.~M. Dickson, J. Phys. Chem. B {\bf 107},  7387  (2003).

\bibitem{Berndt2010}
R. Berndt, J. Kr\"oger, N. N\'eel, and G. Schull, Phys. Chem. Chem. Phys. {\bf
  12},  1022  (2010).

\bibitem{Qiu2003}
X.~H. Qiu, G.~V. Nazin, and W. Ho, Science {\bf 299},  542  (2003).

\bibitem{Dong2004}
Z.-C. Dong, X.-L. Guo, A.~S. Trifonov, P.~S. Dorozhkin, K. Miki, K. Kimura, S.
  Yokoyama, and S. Mashiko, Phys. Rev. Lett. {\bf 92},  086801  (2004).

\bibitem{ifmmode2005}
E. \'{C}avar, M.-C. Bl\"um, M. Pivetta, F. Patthey, M. Chergui, and W.-D.
  Schneider, Phys. Rev. Lett. {\bf 95},  196102  (2005).

\bibitem{Kabakchiev2010}
A. Kabakchiev, K. Kuhnke, T. Lutz, and K. Kern, ChemPhysChem {\bf 11},  3412
  (2010).

\bibitem{Chen2010}
C. Chen, P. Chu, C.~A. Bobisch, D.~L. Mills, and W. Ho, Phys. Rev. Lett. {\bf
  105},  217402  (2010).

\bibitem{Schull2009}
G. Schull, N. N\'eel, P. Johansson, and R. Berndt, Phys. Rev. Lett. {\bf 102},
  057401  (2009).

\bibitem{Schneider2010}
N.~L. Schneider, G. Schull, and R. Berndt, Phys. Rev. Lett. {\bf 105},  026601
  (2010).

\bibitem{Schneider2012}
N.~L. Schneider, J.~T. L\"u, M. Brandbyge, and R. Berndt, Phys. Rev. Lett. {\bf
  109},  186601  (2012).

\bibitem{Avouris1984}
P. Avouris and B.~N.~J. Persson, J. Phys. Chem. {\bf 88},  837  (1984).

\bibitem{Hoffmann2002}
G. Hoffmann, L. Libioulle, and R. Berndt, Phys. Rev. B {\bf 65},  212107
  (2002).

\bibitem{Rossel2010}
F. Rossel, M. Pivetta, and W.-D. Schneider, Surf. Sci. Rep. {\bf 65},  129
  (2010).

\bibitem{Lafferentz2009}
L. Lafferentz, F. Ample, H. Yu, S. Hecht, C. Joachim, and L. Grill, Science
  {\bf 323},  1193  (2009).

\bibitem{Buker2002}
J. Buker and G. Kirczenow, Phys. Rev. B {\bf 66},  245306  (2002).

\bibitem{Galperin2006}
M. Galperin and A. Nitzan, J. Chem. Phys. {\bf 124},  234709  (2006).

\bibitem{Keizer2009}
J.~G. Keizer, J.~K. Garleff, and P.~M. Koenraad, Rev. Sci. Instrum. {\bf 80},
  123704  (2009).

\bibitem{Reecht2013}
G. Reecht, H. Bulou, F. Scheurer, V. Speisser, B. Carri\`ere, F. Mathevet, and
  G. Schull, Phys. Rev. Lett. {\bf 110},  056802  (2013).

\bibitem{Koch2012}
M. Koch, F. Ample, C. Joachim, and L. Grill, Nat. Nanotechnol. {\bf 11},  713
  (2012).

\bibitem{Magoga1997}
M. Magoga and C. Joachim, Phys. Rev. B {\bf 56},  4722  (1997).

\bibitem{supp}
See supplementary material.

\bibitem{Cuevas2010}
J. Cuevas and E. Scheer, {\em Molecular Electronics: An Introduction to Theory
  and Experiment} (World Scientific, Singapore, 2010).

\bibitem{Mujica1994}
V. Mujica, M. Kemp, and M.~A. Ratner, J. Chem. Phys {\bf 101},  6856  (1994).

\bibitem{Schneider2011}
N.~L. Schneider, F. Matino, G. Schull, S. Gabutti, M. Mayor, and R. Berndt,
  Phys. Rev. B {\bf 84},  153403  (2011).

\bibitem{note2}
Possible excitonic effects are not considered in this model.

\bibitem{Rumbles1996}
G. Rumbles, I. Samuel, L. Magnani, K. Murray, A. DeMello, B. Crystall, S.
  Moratti, B. Stone, A. Holmes, and R. Friend, Synth. Met. {\bf 76},  47
  (1996).

\bibitem{Bolognesi2000}
A. Bolognesi, C. Botta, and L. Cecchinato, Synth. Met. {\bf 111-112},  187
  (2000).

\bibitem{Korovyanko2001}
O.~J. Korovyanko, R. \"Osterbacka, X.~M. Jiang, Z.~V. Vardeny, and R.~A.~J.
  Janssen, Phys. Rev. B {\bf 64},  235122  (2001).

\bibitem{Xu1993}
B. Xu and S. Holdcroft, Macromolecules {\bf 26},  4457  (1993).

\bibitem{Salomon2009}
A. Salomon, C. Genet, and T. Ebbesen, Angew. Chem. Int. Ed. {\bf 48},  8748
  (2009).

\bibitem{Dong2010}
Z.~C. Dong, X.~L. Zhang, H.~Y. Gao, Y. Luo, C. Zhang, L.~G. Chen, R. Zhang,
  X.~T.~Y. Zhang, J.~L. Yang, and J.~G. Hou, Nat. Photon. {\bf 4},  50  (2010).

\bibitem{Wu2008}
S.~W. Wu, G.~V. Nazin, and W. Ho, Phys. Rev. B {\bf 77},  205430  (2008).

\bibitem{note3}
The nearly parallel orientation of the wire transition dipole moment (i.e.
  along the long axis of the wire) with the electromagnetic field of the
  excited plasmon mode (i.e. along the tip axis) leads to a much larger plasmon
  amplification than the orthogonal configurations \cite{Thomas2004}. This
  effect is balanced by the reduced amplification expected for large
  tip--sample distances.

\bibitem{Thomas2004}
M. Thomas, J.-J. Greffet, R. Carminati, and J.~R. Arias-Gonzalez, Appl. Phys.
  Lett. {\bf 85},  3863  (2004).

\end{thebibliography}

\end{document}